\nopagenumbers
\def \half {{\textstyle {1\over 2}}}
\def \twothird {{\textstyle {2\over 3}}}
\def \a {{\alpha }}
\def \m {\mu}
\def \n {\nu}
\def \Tr {\rm Tr}
\def \H {{\cal H}}
\def \e {{\epsilon}}
\magnification =\magstep 1
\overfullrule=0pt
\hsize = 6.5truein
\vsize = 8.5truein
\null
\vskip-1truecm
\rightline{IC/97/188}
\rightline{CCNY-HEP 97/8}
\vskip1truecm
\centerline{
United Nations Educational Scientific and Cultural Organization}
\centerline{and}
\centerline{International Atomic Energy Agency} \medskip
\centerline{
INTERNATIONAL CENTRE FOR THEORETICAL PHYSICS}
\vskip1.5truecm
\centerline{\bf Solitons in a six-dimensional super Yang-Mills-tensor
system and non-critical strings }
\vskip 2cm
\centerline{
V.P. Nair *  and S. Randjbar--Daemi}
\footnote{}{* Permanent address: City College of the CUNY, New York, NY 10031.}
\vskip .1 in
\centerline{
International Centre for Theoretical Physics, Trieste, Italy}
\vskip 1.5cm
\centerline{ABSTRACT}
\baselineskip=18pt
\bigskip
In this letter we study a coupled system of six-dimensional $N=1$  tensor
and super Yang-Mills multiplets. We
identify some of the solitonic
states of this system which exhibit stringy behaviour in six dimensions.
A discussion of the supercharges and energy for the tensor
multiplet as well as zero modes is also given. We speculate about the
possible
relationship between
our solution and what is known as tensionless strings.
\vskip 1.5cm
\centerline{
MIRAMARE -- TRIESTE}
\centerline{November 1997}
\vfill\eject
\footline{\hss\tenrm\folio\hss}
\pageno=2
\noindent{\bf 1. Introduction}.

Six-dimensional supersymmetric systems emerge in
many recent studies of
superstring theories. They can
be obtained either by compactifying a ten-dimensional theory or can be
constructed as anomaly-free models in six dimensions. In this note we shall
consider a coupled system of $N=1$ supersymmetric tensor and Yang-Mills
multiplets and study the properties of some of its supersymmetric solitonic
configurations. The zero modes of the
solitonic states constructed in this letter behave like scalars and
spinors in a two-dimensional
subspace of the six-dimensional spacetime and can be grouped
together as the transverse bosonic and fermionic
coordinates of a set of strings propagating in a six-dimensional
spacetime.
There are some intriguing features of our solution
which are suggestive of the tensionless
strings.  We shall
also construct supercharges which generate
the correct supersymmetry transformation of the fields in the tensor
multiplets. We shall then give a Hamiltonian which correctly yields the
quantum equations of motion and show that it vanishes on the subset of
static solitonic backgrounds constructed in this letter. Finally we shall
speculate on possible connections with the non-critical $E_8$ strings.

\noindent{\bf 2.  The Model}

The introduction of tensor multiplets is called for by the requirement of
anomaly cancellation, namely, the
presence of tensor fields in six dimensions enable us to make use of the
Green-Schwarz anomaly cancellation mechanism [1].
The tensor multiplet by itself does not have nonsingular solitonic
solutions. This is one more reason for
considering the tensor multiplet coupled to Yang-Mills fields.

The Yang-Mills multiplet consists of the gauge field
$A_\mu$, the gaugino
$\lambda^i$ and an auxilary field $Y^{ij}$. The tensor multiplet involves a
scalar $\sigma$, a second rank antisymmetric tensor $B_{ab}$ and a tensorino
$\chi^i$. All of the fermions are chiral. The index $i$ is an $SU(2)$-index
and the fermions satisfy a symplectic
Majorana condition. The supersymmetry parameters will be taken to have
positive six-dimensional chirality.

The supersymmetry transformations of the fields in our system are as
follows [2].
$$\eqalign{
\delta A_a&= -\bar\epsilon~\Gamma_a\lambda\cr \delta \lambda^i &=
{1\over 8}\Gamma^{ab} F_{ab} \epsilon^i - {1\over 2} Y^{ij}\epsilon_j\cr
\delta Y^{ij} &= - {\bar\epsilon}^{(i}~\Gamma^a D_a\lambda^{j)}\cr} \eqno(1)
$$
The corresponding rules for the  tensor multiplet
coupled to Yang--Mills are given by
$$\eqalign{
\delta \sigma &= \bar\epsilon\chi\cr
\delta\chi^i &= {1\over
48}\Gamma^{abc} {H}^{+}_{abc}
\epsilon^i +{1\over 4}\Gamma^a\partial_a\sigma\epsilon^i - {\a'\over 4}{\rm
Tr}\,\Gamma^a\lambda^i
\bar\epsilon\Gamma_a\lambda \cr
\delta B_{ab} &= -\bar\epsilon\Gamma_{ab}\chi - \a' {\rm Tr}\,
A_{[a}\bar\epsilon
\Gamma_{b]}\lambda
\cr}\eqno(2)
$$
where
$$\eqalign{
{ H}_{abc} &= 3\partial_{[a} B_{bc]} + 3\a'{\rm Tr} \left(A_{[a}\partial_b
A_{c]}+{\twothird} A_a A_b A_c\right)\cr
{ H}_{abc}^{\pm} &= {1\over 2}\biggl ( { H}_{abc} \pm \tilde {
H}_{abc}\biggr ).\cr}
$$

The closure of the supersymmetry algebra leads to the following field
equations for various fields,
$$\eqalignno{
{ H}_{abc}^{-} &= -{\a'\over 2}{\rm Tr}\,( \bar\lambda
\Gamma_{abc}\lambda)&(4a)\cr
\Gamma^a\partial_a\chi^i &= \a' {\rm Tr}\,\left({1\over 4}\Gamma^{ab}
F_{ab} \lambda^i + Y^{ij}\lambda_j\right)&(4b)\cr \partial^2 \sigma &= \a'
{\rm Tr}\,\left(-{1\over 4}F^{ab}F_{ab} - 2\bar\lambda \Gamma^a
D_a\lambda + Y^{ij}Y_{ij}\right).&(4c)\cr }$$
Further, by virtue of its definition,
$H_{abc}$ satisfies the identity
$$
\partial_{[a}{ H}_{bcd]}^+ = \a' {\rm tr}\,\left({3\over 4} F_{[ab}F_{cd]}
-\bar\lambda\Gamma_{[abc}D_{d]}\lambda\right) \eqno(5)
$$

\noindent{\bf 3. The Solution}.

We shall look for a bosonic background configuration
in which all the
fermions as well as the auxiliary
field $Y^{ij}$ will vanish. The six-dimensional coordinates will be chosen
as $x^+$, $x^-$ and $x^{\m}$ where $\m =1,...4$.
We shall consider a multi instanton-type configuration in the ${\bf R}^4$
spanned by $x^{\m}$. We shall show
that the moduli of this instanton can depend on $x^+$. This will require
that the $A_+$-component of the vector potential is different from zero. In
this sense our solution looks like a static monopole configuration in
the six-dimensional spacetime in which $x^-$ is taken to be the time
coordinate. This configuration will
preserve half the six-dimensional $N=1$ supersymmetry.

It follows from (4a) that if $\lambda =0$, then $H$ is selfdual.\footnote
{*}{ We shall henceforth drop the superscript $+$ from H}. Now
setting $\delta \lambda$ and $\delta\chi$
equal to zero we obtain
$$
\Gamma^{ab} F_{ab} \epsilon =0,~~~~~~~~~ \left( \Gamma^a\partial_a \sigma
+{1\over 12} \Gamma^{abc}H_{abc} \right ) \epsilon =0
\eqno(6)
$$
To satisfy these equations, we can choose
$\epsilon=\left(\matrix{\varepsilon\cr 0}\right)$, where,
 $\gamma_5 \varepsilon =\pm
\varepsilon$, and
 $\gamma_5$ gives the four-dimensional chirality. We shall first discuss
the case
of positive chirality; the case of negative chirality can be obtained by
essentially
trivial change of some selfduality conditions. With this choice the fields
must obey the equations $$
\eqalignno{
H_{05\m} &= -\partial_\m \sigma &(7a)\cr H_{0\m \n} &= {\tilde H_{0\m\n}}=
\half \e_{\m\n \alpha \beta}H_{0\alpha \beta}&(7b)\cr F_{\m\n}&= {\tilde
F_{\m\n}}=
\half \e_{\m \n \alpha \beta}F_{\alpha \beta}&(7c)\cr }
$$
together with $F_{+-}=0, ~F_{-\m}=0, ~\partial_- \sigma =0$. Choosing the
guage $A_- =0$
, these reduce to
$\partial_- A_\m = \partial_- A_+ = \partial_- \sigma =0$.

The constraint (5) for $H_{abc}$,  expressing its coupling to the Yang-Mills
fields via the Chern-Simons 3-form, now gives
the following conditions,
$$\eqalignno{
\partial_- H_{+\m\n}&=0&(8a)\cr
\partial_\lambda H_{+\lambda \alpha} &= \partial_+ \partial_\alpha \sigma ~
-2c \Tr (F_{\lambda \alpha}{ F}_{+\lambda})&(8b)\cr \partial_\m \partial_\m
\sigma
 &= -{c\over 2} \Tr (F_{\m\n} {\tilde F}_{\m\n})&(8c)\cr
}
$$
where $c= 3 \a '/4$.
Further, $H_{-\m\n}=0$ and $H_{+\m\n}= {\tilde H}_{+\m\n}$. Setting the
auxiliary field
$Y^{ij}$ to zero implies $D_a F^{ab}=0$ [2]. The only nontrivial surviving
component of this equation is $$
D_\lambda (D_\lambda A_+ - \partial_+ A_\lambda )=0 \eqno(9)
$$
where $D_\lambda A_+ = \partial_\lambda A_+ + [A_\lambda ,A_+]$.

The strategy for solving these equations is as follows. We first choose
$F_{\m\n}$ to be a multi-instanton configuration in ${\bf R}^4$.
Then equation (8c) gives $\sigma$, and (7a) gives $H_{05\m}$. Since
$D_\lambda D_\lambda$ is
invertible in the instanton background, (9) can be uniquely solved for $A_+$.
 Finally, equation (8b) can be solved, consistently with its selfduality,
to get $H_{+\m\n}$.
 As a consequence of $\partial_- A_\m = \partial_- A_+ =0$, the instanton
parameters,
collectively denoted  by $\xi$, obey
the condition $\partial_- \xi =0$, but they can, of course, depend on $x^+$.
(They are thus left-moving modes in the $(x^0,x^5)$-subspace.)

Using the selfduality of $H_{+\m\n}$, we can rewrite (8b) as $$
\partial_\lambda \partial_\lambda H_{+\m\n} = (\partial_\m J_\n
-\partial_\n J_\m ) +\half \epsilon_{\m\n \alpha \beta}
(\partial_\alpha J_\beta -\partial_\beta J_\alpha ) \eqno(10)
$$
where $J_\alpha = \partial_+ (\partial_\alpha \sigma )- 2c \Tr ( F_{\lambda
\alpha }F_{+\lambda})$. It is easy to
 see that $\partial_\alpha J_\alpha =0$, as required by the consistency of
the equations. Since the four-dimensinal
Laplacian is invertible, the above equation can be easily solved, once we
have $J_\alpha$. For gauge group
$SU(2)$, $A_+$ is given by $$
A_+^a = \int d^4y~ \Delta^{ab}(x,y) \epsilon^{bkl} (A^k_\lambda \partial_+
A^l_\lambda )(y)\eqno(11)
$$
where the Green's function $\Delta^{ab}(x,y)$ for $D_\lambda D_\lambda$ in
the instanton bcakground is given in reference [3]. To make the above
solutions explicit, we can, for example, take the 't Hooft ansatz for
instantons, viz.,
$A^a_\m = {\bar \eta}^a_{\m\n} \partial_\n (\log \phi )$ where $\phi = 1 +
\sum_1^N {\rho_i^2 / (x -a_i)^2} $ and insert it in various equations
above. In this case, $\sigma$, for example, becomes $
2c  \partial_\m \phi \partial_\m \phi /\phi^2 $.

\noindent{\bf 4. Supercharges.}

The condition that half of the supersymmetry is
unbroken leads to BPS-like
constraints.
For the theory we are considering, there is no known action or Hamiltonian
formulation without introducing additional degrees of freedom. For the free
tensor multiplet,
 possible  Hamiltonian formulations have been given in [4, 5]. It is
posssible to extend the formulation of [4] to the
 case that the Yang-Mills fields are treated as background fields.
Specifically, we can choose $\epsilon^1 = (\alpha ,0),~ \chi^1 = (0,v)$
where $\alpha, ~v$ are four-component
spinors, with $\e^2, \chi^2$ given by the symplectic Majorana condition.
The supersymmetry variation for the field $v$ is then given as $$
\delta v = \left[ -{1\over 8} \gamma^a \gamma^b H^*_{ab} + {1\over 4} (-
\pi + \gamma^a \partial_a \sigma ) \right] \alpha \eqno(12) $$
where $H^*_{ab} = (1/6) \e_{abcde}H^{cde},~ \pi= \partial_0 \sigma.$ The
indices $a,b$ now range from $1$ to $5$. The bosonic fields obey the
commutation rules $$\eqalign{
[\sigma (x), \pi (y)]&= 4i \delta (x-y)\cr [B_{ab}(x), B_{cd}(y)]&= 4i
\e_{abcde}\partial_e G(x,y)\cr }\eqno(13)
$$
where $G(x,y)$ is the five-dimensional Coulomb Green's function. The
supercharge for the tensor multiplet is given by
$$
Q_r = \int v^\dagger_s \left[ -{1\over 8} \gamma^a \gamma^b H^*_{ab}
+{1\over 4} (-4\pi +\gamma^a \partial_a \sigma )\right]_{sr} \eqno(14)
$$
(This generates the correct transformations in the gauge
$B_{0a}=0,~\partial_a B^{ab}=0$.) The supercharges obey the algebra $\{
Q_r,{\bar Q}_s \}= \half ({\tilde \H}\delta_{rs} +\gamma^a_{rs}{\tilde P}_a)$
with
$$
\eqalign{
{\tilde \H}&= {1\over 8}\int \left[ \pi^2 +(\partial \sigma )^2 +\half
(H^*)^2 \right]\cr
{\tilde P}_a &= -{1\over 8} \int \left[ (\pi \partial_a \sigma +\partial_a
\sigma \pi ) + {1\over 4} \e_{amnpq} H^*_{mn}H^*_{pq} +2 \partial_b \sigma
~H^*_{ba}\right]\cr }\eqno(15)
$$
${\tilde \H}, {\tilde P}_a$ are the Hamiltonian and the momentum for the
bosonic fields of the $free$ tensor multiplet. With a Yang-Mills
background, one cannot expect a super-Poincar\'e algebra and one cannot
read off the Hamiltonian from this algebra. Rather, the Hamiltonian, for
the bosonic fields, is now given by $$
\H = {\tilde \H} +\int\left[ \sigma J
- {c\over 8} \e_{mnpqr}\omega_{0mn}(\partial_r B_{pq})\right] \eqno(16)
$$
where $J= -(\a '/4){\Tr} \left( -{1\over 4}F^{ab}F_{ab}-2{\bar\lambda}
\Gamma^a D_a\lambda +Y^{ij}Y_{ij}\right)$ and $H^*$ entering $\H$ now does
contain the Yang-Mills Chern-Simons contribution. The condition of
half-supersymmetry gives the saturation condition ${\tilde \H}= \vert
{\tilde P}\vert$, where, for our solution in the static limit,
$\vert {\tilde P}\vert = (1/4)\int (\partial \sigma )^2$.
Using the equation of motion for $\sigma$, we then find $\H = 0$
for our solution in the static limit. Of course, the Hamiltonian (16)
treats the gauge fields as a background. One cannot simply add the
Yang-Mills Hamiltonian to this since the equations of motion for the latter
are unaffected by the tensor multiplet. Nevertheless, the fact that $\H =0$
is indicative of a possible  connection to the tensionless string, for
which we shall give more evidence in
the next section.

\noindent{\bf 5. String Interpretation.}

To see the stringy interpretation of our
solution, we need to analyze its
moduli or zero mode structure.
 From the above equations, we see that, given the gauge field $F_{\m\n}$,
all the fields are uniquely
determined upto the addition of the freely propagating six-dimensional
waves for the tensor multiplet.
\footnote{*}{ Note the soliton
does not modify their propagation.}
Therefore only the zero modes correspond to the moduli of the instantons.

In order for our models to be mathematically meaningful they should be free
from local and global gauge anomalies. In the absence of hypermatter, the
 gauge groups  $SU(2),~SU(3),~G_2,~F_4,~E_6,~E_7$ and
$E_8$ can be made perturbatively anomaly-free with the help of the
Green-Schwarz prescription [1]. However, since the homotopy
group $\Pi_6$ of
the first three groups in this list are nontrivial, these theories will
harbour global gauge anomalies [6]. To
make them consistent we need to introduce hypermatter for these theories[7].
The allowed matter contents for the
cancellation of the global [7] as well as the local [8] anomalies in the
presence of one tensor multiplet are
$n_2=4, 10$
for $SU(2)$, $n_3=0, 6, 12$ for $SU(3)$ and $n_7=1, 4, 7$ for $G_2$, where
$n_2, n_3$ and $n_7$ represent the number of the doublets
for $SU(2)$, triplets for $SU(3)$ and 7-dimensional representation of $G_7$,
respectively. All other gauge groups are free from
global anomalies and they can be made free from perturbative anomalies
(using the Green-Schwarz prescription)
if appropriate amount of hypermultiplets are
taken together with the gauge and the tensor multiplets [7, 8].

For the gauge group
$SU(2)$, for the four-dimensional space being ${\bf R}^4$ and for instanton
number $k$,
we have $8k$ bosonic moduli corresponding to the instanton positions,
scale sizes and group orientations.
(The equations of motion, despite the appearance of the dimensional
parameter $c$,
have scale invariance and give the scale size parameter in the solutions.)
These moduli appear
in the solution for the fields $B_{ab}$ as well.

The surviving supersymmetry has $\gamma_5 \varepsilon =\varepsilon$, i.e.,
left-chirality in the four-dimensional sense
corresponding to a $(4,0)$ world-sheet supersymmetry for the solitonic
string. There must necessarily be fermionic zero modes.
For the gauginos, we have $4k$ zero modes for the gauge group $SU(2)$,
which are
of right-chirality in the four-dimensional sense and are in the
right-moving sector.
The Dirac equation for the gauginos along with the half-supersymmetry condition
shows that the gaugino zero mode parameters are constants; the bosonic
parameters are constant as well, by supersymmetry.
The fermionic zero mode parameters are complex, i.e.,
we have $8k$ real Grassman parameters which balance the $8k$ bosonic
parameters.
Some of the fermionic zero
modes correspond to the supersymmetries which are broken by the background
and can be obtained by such supersymmetry variations. With hypermatter, there
are also hyperino zero modes, which are in the left-moving sector. There is no
supersymmetry for these modes and generically there are no hyperscalar zero
modes.

For higher gauge groups, there will be more moduli. Thus, for example, for
$SU(3)$, with the standard
embedding of the instanton and $n_3=0$, we have $12k$ bosonic parameters
and $6k$ fermionic parameters. It is easy to see that
the number of moduli for all of the anomaly-free gauge groups listed above
is always a multiple of $4$. We may thus interpret these solutions as
six-dimensional strings
with $4$ tranverse coordinates  corresponding to the zero modes for the
broken translational symmetries. The remaining
zero modes can be regarded as additional world-sheet degrees of freedom.
In this way for instanton number
$k$, we have $k$ strings with $(4,0)$ world-sheet supersymmetry.

As an example, consider an $SU(2)$ theory with 10 hypermatter doublets [9]
. In this case,
 for instanton number equal to one, we have eight
instanton moduli, eight gaugino zero modes for the right-moving sector and
20 hypermatter zero modes for the left-moving sector.
 The $SU(2)$ symmetry can be spontaneously broken  by  vacuum expectation
values of
 the scalars originating from the moduli corresponding to 
the global $SU(2)$
rotations and the scale size of the instanton.
 By supersymmetry this should remove four of the gaugino zero modes from the
right moving sector by giving them a non zero mass, which will also eat up
four hyperino zero modes in the left moving sector.
 One is left with four moduli for
the instanton, four gaugino modes in the right-moving sector and 16
hyperino zero
modes in the  left-moving sector. These 16  hyperino zero modes presumably
generate a left moving $E_8$ current algebra.
This looks like the spectrum of the non critical string which lives in the
boundary of a membrane joining
a $5$-brane to a $9$-brane in $M$-theory and which becomes tensionless as
the $5$-brane approaches the $9$-brane [10]. It has been
argued in [11] that the same model corresponds to one of the phases of the
$F$-theory.

There are also independent solutions with the opposite chirality.
The choice $\gamma_5 \varepsilon =- \varepsilon$ leads to antiselfdual
$H_{+\m\n},~F_{\m\n}$ with $A_+=0$ and $\partial_+ \xi =0$.

The solution we have
obtained is a static one.
The choice of four-dimensional chirality as $\gamma_5 \epsilon =\pm
\epsilon$ leads to static solitons.
By Lorentz boosts, it is possible to obtain a
solution whose center of mass is moving at a constant velocity.
For a moving soliton, the condition $\gamma_5 \epsilon =\pm
\epsilon$ must be modified.
Consider, for example, the one-soliton (one-instanton)
solution. We choose the supersymmetry parameters $\varepsilon$ as
$S\varepsilon_{(0)}$ where $S=\exp ( -\half \omega^\mu \gamma_\mu)
\approx 1 -\half \omega^\mu \gamma_\mu$ and $\varepsilon_{(0)}$ obeys $\gamma_5
\varepsilon_{(0)}=\varepsilon_{(0)}$. (For small velocities, the parameter
$\omega^\mu \approx v^\mu$, the velocity.)
The vanishing of the gaugino variation,
viz., $\Gamma^{ab}F_{ab}\epsilon =0$, now gives,
to first order in $v^\mu$,
$$
\eqalign{
F_{\mu\nu} - {\tilde F}_{\mu\nu}&=0\cr
F_{-\mu}+ {1\over {\sqrt {2}}}F_{\mu\nu}v^\nu &=0\cr
F_{+-} - {1\over {\sqrt{2}}} F_{+\nu}v^\nu &=0\cr
}\eqno(17)
$$
To this order, $F_{\mu\nu}$ is still selfdual. The other two equations are
seen to be satisfied if we take the instanton position $a^\alpha$ to
move with velocity $v^\alpha$, i.e., $\partial_0 a^\alpha =v^\alpha$.
(We can make a gauge transformation $A_{-}\rightarrow A_{-}-
(1/{\sqrt{2}})A_\mu v^\mu $ to restore the $A_{-}=0$ gauge.) There is a
similar set of statements for the vanishing of the tensorino variation.
What we have shown is that a soliton whose center of mass is moving at a
constant
velocity $v^\alpha$ is also a supersymmetric solution with supersymmetry
parameters being $S\varepsilon_{(0)},~\varepsilon_{(0)}$ having definite
four-dimensional chirality.
\vskip .2in
After completion of this paper there appeared ref. [12] which overlaps with
our section 3.
\bigskip
\noindent{\bf Acknowledgments.}
We are grateful to E. Gava, K.S. Narain, E. Sezgin
and G. Thompson for useful discussions.
We are particularly indebted to C. Vafa for many helpful discussions, for
drawing our attention to
references 7 and 8 and his criticisms of our earlier treatment of the zero
modes.
VPN also thanks J.T. Liu for useful comments.

The work of VPN was supported in part by NSF Grant Number PHY-9322591.
\vskip .2in
\noindent{\bf References}
\item{1.} M. Green, and J. Schwarz,  Phys. Lett. {\bf B149} (1984) 117;
S. Randjbar-Daemi, Abdus Salam, E. Sezgin and J. Strathdee,  Phys. Lett.
{\bf B151} (1985) 351;
J. Schwarz, Phys. Lett. {\bf B371} (1996) 223, hep-th 9512053;
N. Seiberg,  {\bf B 390} (1997) 169,  hep-th 9609161
\item{2.} E.Bergshoeff, E.Sezgin and E.Sokatchev, Class. Quantum Grav. {\bf
13} (1996) 2875.
\item{3.} L.S. Brown {\it et al}, Phys. Rev. {\bf D17} (1978) 1583.
\item{4.} I. Giannakis and V.P. Nair, Phys. Lett. {\bf B409} (1997) 145,
hep-th 9702024.
\item{5.} M. Perry and J. Shwarz, Nucl. Phys. {\bf B489} (1997) 47;  P.
Pasti, D. Sorokin and M. Tonin, Phys. Rev. {\bf D55} (1997) 6292; Phys.
Lett. {\bf B398} (1997) 41.
\item{6.} E. Witten, Phys. Lett. {\bf B117} (1982) 324: S. Elitzur and V.P.
Nair Nucl.Phys. {\bf B243}(1984) 205.
\item{7.} M. Bershadsky and C. Vafa, hep-th 97030167.
\item{8.} U. H. Danielsson, G. Ferretti, J. Kalkkinen and P. Stjernberg,
hep-th 9703098.
\item{9.}P. Mayr and C. Vafa, unpublished.
\item{10.}Ori J. Ganor and Amihay Hanany hep-th 96020120.
\item{11.} A. Klemm, P. Mayr and C. Vafa, hep-th 96070139.
\item{12.}J. Duff, J.T. Liu, H. Lu, C.N. Pope.  hep-th  hep-th/9711089.

\end